\documentclass[pre,twocolumn,superscriptaddress,floatfix]{revtex4}
\pdfoutput=1
\usepackage{epsfig}
\usepackage{float}
\usepackage{amsmath}
\usepackage{graphicx}
\usepackage{amssymb}
\usepackage{mathtools}
\usepackage{color}

\begin{document}

\title{Evaluating the Laplace pressure of water nanodroplets from simulations}

\author{Shahrazad M.~A.~Malek}
\affiliation{Department of Physics and Physical Oceanography,
Memorial University of Newfoundland, St.~John's, NL, A1B 3X7, Canada}

\author{Francesco Sciortino}
\affiliation{Dipartimento di Fisica, Sapienza Universit\`a di Roma, Piazzale Aldo Moro 5, 00185 Roma, Italy}

\author{Peter H.~Poole}
\affiliation{Department of Physics, St.~Francis Xavier University, Antigonish, NS, B2G 2W5, Canada}

\author{Ivan Saika-Voivod}
\affiliation{Department of Physics and Physical Oceanography,
Memorial University of Newfoundland, St.~John's, NL, A1B 3X7, Canada}

\date{\today}

\begin{abstract}

We calculate the components of the microscopic pressure tensor as a function of radial distance $r$ from the centre of a spherical water droplet, modelled using the TIP4P/2005 potential.
To do so, we modify a coarse-graining method for calculating the microscopic pressure [T.~Ikeshoji, B.~Hafskjold, and H.~Furuholt, Mol.~Simul.~{\bf 29}, 101 (2003)] 
in order to apply it to a rigid molecular model of water.  As test cases, we study nanodroplets  ranging in size from 776 to 2880 molecules at 220~K.
Beneath a surface region comprising approximately two molecular layers, the pressure tensor becomes approximately isotropic and constant with $r$.  We find that the dependence of the pressure on droplet radius is that expected from the Young-Laplace equation, despite the small size of the droplets.

\end{abstract}

\maketitle

\section{Introduction}


Small droplets of liquid water are important to atmospheric science and technological applications, and understanding the properties and role of the surface is increasingly important as droplets become nanoscopic.  Surface effects can profoundly influence the mechanism and rate of crystallization in general.  In water, the role of surface freezing is still unresolved~\cite{Akbari2017}.

Significant to much of the discussion is the Laplace pressure, the pressure difference between the interior and exterior of a droplet of radius $R$ arising from the liquid-vapour surface tension $\gamma$, as quantified by the Young-Laplace equation for droplets,
\begin{equation}\label{eq:laplace}
\Delta P = \frac{2 \gamma}{R}.
\end{equation}
Galli and coworkers modelled the effect within nanodroplets of the Laplace pressure on nucleation rates~\cite{Galli2013}.  They argued that since the interior of the nanodroplet is at a higher pressure, the liquid there is less supercooled on account of the decreasing melting temperature of ice Ih with increasing pressure. Hence nucleation rates should be greatly diminished in the interior.   Espinosa et al~\cite{CHANTAL} went on to show that the liquid-Ih surface tension also increases with increasing pressure, further suppressing nucleation.   By contrast, the nanodroplet surface, though prone to disorder, experiences a negative pressure, and should thus be more supercooled and therefore enhance nucleation rates.  The simulations of Ref.~\cite{Galli2013} showed that nucleation rates for mW~\cite{Moore2009} water nanodroplets are progressively and greatly suppressed as nanodroplet size decreases, and that the rates are the same within error for $R\ge3.1$~nm  when compared to the bulk at the same density.  For smaller nanodroplets, the difference in rates between droplets and bulk at the same density is significant.  The authors argue, however, that for real water, for which the density difference between liquid and crystal at melting is larger than in mW water, surface nucleation should be favoured in microdroplets.  We note that while the authors estimated the Laplace pressure through Eq.~\ref{eq:laplace} and provided a check of the equation by determining the pressure of the bulk at the same density as inside the nanodroplets,  they did not explicitly calculate the pressure inside the droplets.  Nor is it clear to what extent Eq.~\ref{eq:laplace} should hold for more realistic models of water, such as the TIP4P model~\cite{TIP4P1983}  and related potentials~\cite{TIP4Ppotentials}.

The insights of Ref.~\cite{Galli2013} have been enriched by the work of Haji-Akbari and Debenedetti~\cite{Debenedetti2017} on water nanofilms.  They found that nucleation rates obtained using the TIP4P/ice~\cite{TIP4Pice2006}  model of water are enhanced by a factor of $10^7$ within the nanofilm in comparison to the bulk.  The enhancement stems not from the interface, where crystal-like ordering is reduced, but rather from a relative abundance of ``double-diamond cages'' over hexagonal cages in the interior of the film compared to bulk.  The latter cage type is less favourable for nucleation. Their work therefore indicates the importance of subtle changes in structure arising from the finite extent of the system, and diminishes the importance of the negative pressure near the interface.  However, this study was conducted on films,  where the internal pressure is no different from the ambient, and therefore did not address the role of the Laplace pressure on the interior.

Recent experiments on microdroplets, for which the Laplace pressure is likely negligible, have pushed the limits of observing liquid water below the bulk homogeneous nucleation limit of 235~K by determining nucleation rates down to 227~K~\cite{Nilsson2014}.  Nucleation rates at significantly lower temperatures have been measured for nanodroplets with radii of just a few nanometers~\cite{Bartell1995,Wyslouzil2012}, for which the Laplace pressure is likely significant.  An experimental study of water clusters in the range of 100-1000 molecules showed that crystallization may be entirely suppressed below roughly 275 molecules~\cite{Buck2012}, at which point surface effects may dominate and the Laplace pressure would be quite high.
Given  that experiments  probe ever smaller systems, it is crucial to develop a better understanding of the basic physical properties of nanodroplets, including the pressure.

The theoretical and experimental developments described above all point to the need for a detailed analysis of the microscopic pressure tensor within water nanodroplets and its connection to the Laplace pressure.  This is  the subject of this paper.  
The work on ST2 water clusters of Brodskaya et al.~\cite{brodskaya1987,brodskaya1993} found significantly elevated pressures within nanodroplet interiors.  Thompson et al.~\cite{thomp} provided a detailed description of the methodology for calculating the pressure tensor in droplets in the context of Lennard-Jones particles. 
We base our calculations on the work by Ikeshoji et al.~\cite{FURUHOL}, who developed a coarse-grained scheme for calculating the molecular-scale pressure for simple particles interacting with radial potentials.  The advantages of their method include improved statistics over non-coarse-grained methods (e.g.~\cite{thomp}), as well as the ability to directly calculate both the normal and transverse components of the pressure tensor.  
The method was applied to a molecular model of water, SPC/E~\cite{Straatsma1987}, in a study of methane hydrate droplets embedded in ice~\cite{subbotin2007}, but no details on how the method was modified for molecules were given. 
The method of Ref.~\cite{FURUHOL} was later generalized to molecules in a way that considered multibody intramolecular interactions, and applied to non-rigid chain-like organic molecules interacting with a coarse grained-model for water~\cite{ikeshoji2011}.
However, for rigid multi-site water models such as TIP4P/2005, it is more straightforward to modify Ref.~\cite{FURUHOL} in a way that does not require the 
consideration of intramolecular interactions, i.e., forces of constraint.
It is this latter approach that we present here.  That is, we adapt the method of Ref.~\cite{FURUHOL} to 
TIP4P/2005 water nanodroplets, and give details of the calculation.

This paper is organized as follows.  Section~\ref{sec:sim} describes our molecular dynamic simulations of TIP4P/2005 water nanodroplets.  In Section~\ref{sec:press} we show in detail how we adapt and apply the method introduced in Ref.~\cite{FURUHOL} to water, comment on the utility of the method in terms of independently calculating the normal and transverse components of the local pressure tensor, and introduce an energy-based approximate method of calculating the local isotropic pressure and use it as a check of our results.  We present the pressure components as functions of radial distance from the centre of mass of a nanodroplet and validate the form of Eq.~\ref{eq:laplace} in Section~\ref{sec:res}, before concluding in Section~\ref{sec:dis}.

\section{Simulations}\label{sec:sim}

We simulate nanodroplets of $N= 776$, 1100, 1440 and 2880 water molecules interacting through the TIP4P/2005 water model~\cite{Vega2005}.
All simulations are done at temperature $T=220$~K, where the vapour pressure is negligible.
For $N$ = 1440 and 2880, we initially prepare a droplet system of a given size by placing $N$ water molecules randomly in a rather large cubic simulation box and simulating at constant volume. The molecules naturally condense into a droplet surrounded by a very low density vapour. 
The equilibrated configuration  is then run for many relaxation times to get equilibrium properties of the droplets.
We produce  two  spherical droplets of size $N$ = 776 and 1100 by removing molecules beyond an appropriate radial distance from the centre of an equilibrated $N=1440$ droplet. 
For the $N=776$ system, the simulation box length $L=15$~nm.  For the larger droplets $L$ = 20~nm. 
We use a potential cutoff of $L/2$, and  employ periodic boundary conditions to ensure that vapour molecules can return to the
droplet in order to avoid eventual evaporation.  
The box is large enough to avoid any direct interaction between the water droplet and its periodic images.
With this setup, molecules within the droplet interact through the full, untruncated potential, including electrostatic interactions.
We use Gromacs v4.6.1~\cite{GROMACS} to carry out the molecular dynamics simulations. We hold $T$ constant with the Nos{\'e}-Hoover thermostat. The equations of motion are integrated with the leap-frog algorithm with a time step of 2 fs.
The total simulation times for the four droplet sizes, in order of increasing $N,$ are 862, 633, 593 and 182~ns.

To determine equilibration and relaxation times, we monitor the decay of the bond autocorrelation function $\phi(t)$, which gives the probability that a bond present at time $t=0$ remains unbroken until time $t$~\cite{bonds}.
Two molecules $i$ and $j$ are considered bonded if the distance between their O atoms is less than $0.32$~nm, the location of the first minimum in the oxygen-oxygen radial distribution function of bulk water at ambient conditions.  The calculation of $\phi(t)$ is sensitive to the sampling interval, which in our case falls between 0.2 and 0.8~ns.  We can not discriminate between persistent and reformed bonds on times shorter than our sampling time, and so our $\phi(t)$ provides an upper bound on the true value.

Error bars for various quantities are calculated by taking the standard deviation in a quantity over all sampled equilibrium configurations, and dividing by $\sqrt{n_{\rm ind}}$, where $n_{\rm ind}=t_{\rm eq}/\tau_\phi$ is the estimated number of independent configurations sampled, $t_{\rm eq}$ is the duration of the equilibrated time series used for averaging, and $\tau_\phi$ is the time at which $\phi(t) \le {e}^{-1} \approx 0.368$.  
For example, for the $N=1100$ droplet, the simulation is carried out for a total of  633~ns, the first 129~ns of which are discarded, leaving $t_{\rm eq}=504$~ns.  Our determination of $\phi(t)$ is not very well resolved in time, but we determine that $\phi(0.8~{\rm ns})=0.08$ and so we set  $\tau_\phi=0.8$~ns and hence $n_{\rm ind} \approx 500/0.8 = 625$.  
Our estimates for the number of independent configurations sampled in equilibrium for the other sizes are 1917 ($N=776$), 1588 ($N=1440$) and 48 ($N=2880$).

\section{Microscopic pressure}\label{sec:press}

\subsection{Pressure profiles}

To calculate the normal $P_N(r)$ and tangential $P_T(r)$ components of the pressure tensor as a function of radial distance $r$ from the centre of mass of the water nanodroplet, we follow the prescription of Ikeshoji {\it et al.}~\cite{FURUHOL}  for a spherical geometry.  Below we reproduce their approach, which uses a coarse graining wherein the pressure components at $r$ are calculated as averages over a thin spherical shell of finite thickness 
in order to improve statistics and to avoid divergences in $P_T(r)$.

\begin{figure}[!htb]
 \centering
 \includegraphics[width=1.0\columnwidth]{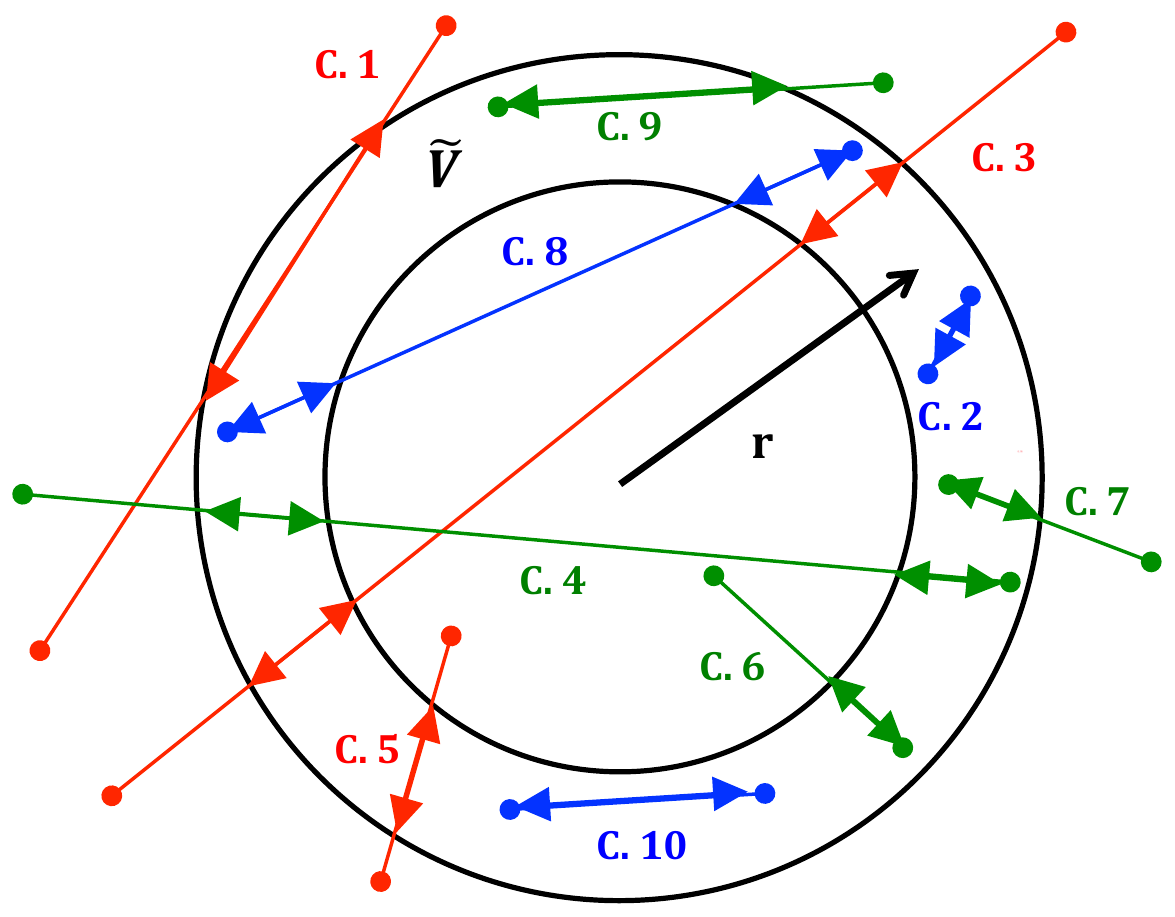}
 \caption[A sketch of all possible contributions to $P$ in $\tilde V$ from the CG method]{
 A sketch of all possible contributions to $P$ in $\tilde V$ from the coarse-graining method of Ikeshoji {\it et al.}~\cite{FURUHOL}. 
 See Table~\ref{tablambda} for details.
 The contours $C_{ij}$ are line segments between molecules $i$ and $j$ (filled circles).
 The portions of $C_{ij}$ between arrows contribute to the pressure.
 $\tilde V$ is a spherical shell of inner radius $R_{\rm in }=r-\Delta R/2$ and outer radius $R_{\rm out}=r+\Delta R/2$.
 }
 \label{cgcases}
 \end{figure}

Their method was presented for particles interacting through central forces.  We introduce adaptations required since the pair force between water molecules is not central (although the site-site interactions are).  The generalization is straightforward since only the intermolecular forces need to be considered and they need not be central~\cite{brodskaya1987,brodskaya1993, tildesley1995}.  In order to present the reader with a self-contained explanation of the method, we have  reproduced relevant portions of Ref.~\cite{FURUHOL} here.
To be more explicit, Eqs.~\ref{PKP} to~\ref{unitv} and their development are adapted from Ref.~\cite{FURUHOL}, albeit with slightly different notation, while Eqs.~\ref{lambdaint1} to~\ref{eq:sigmaT} have been modified because of the non-central force between molecules.  Fig.~\ref{cgcases} is adapted from~\cite{FURUHOL} to explicitly include all types of molecular pair contributions.  We introduce Table~\ref{tablambda} to provide mathematical details that complement Fig.~\ref{cgcases}.  

Schofield and Henderson~\cite{Schofield} showed that the pressure tensor at a point $\bf {R}$ in space is given by~\cite{Schofield,thomp},
\begin{equation}
P^\prime_{\alpha\beta}({\bf R}) = \left<P^\prime_{c,\alpha\beta}({\bf R})\right>+\left<P^\prime_{k,\alpha\beta}({\bf R})\right>,
\label{PKP}
\end{equation}
where
\begin{equation}
\left<P^\prime_{k,\alpha\beta}({\bf R})\right>=k_BT\rho({\bf R}) \delta_{\alpha\beta},
\label{PK}
\end{equation}
is the kinetic part, and follows directly from the local equilibrium density $\rho({\bf R})$.  The brackets $\left< \dots\right>$ indicate an 
ensemble average, i.e., an average over a set of equilibrated configurations, and $\delta_{\alpha\beta}$ is the Kronecker delta.  Pressures annotated with a prime indicate that the pressure is calculated at a single point in space.  Pressures without primes refer to quantities that are coarse-grained (averaged) over a small volume.

The configurational contribution is obtained from intermolecular pair forces, and is given by,
\begin{equation}
P^\prime_{c,\alpha\beta}({\bf R}) = \frac{1}{2}\sum_i\sum_{j\ne i}P^\prime_{ij,\alpha\beta}({\bf R}),\label{pconf}
\end{equation}
where the molecular pair-wise contribution to the pressure is given by,
\begin{equation}
P^\prime_{ij,\alpha\beta}({\bf R}) = \int_{C_{ij}} f_{ij,\alpha} \, \delta({\bf R}-{\bf l}\,\,)dl_\beta\label{pij},
\end{equation}
where $f_{ij,\alpha}$ is the $\alpha$ component of the force on molecule $j$ due to molecule $i$, ${\bf f}_{ij}$,
$\delta({\bf R}-{\bf l})$ is the Dirac delta function, $C_{ij}$ is a contour from $i$ to $j$, ${\bf l}$ is a vector indicating a point on $C_{ij}$, and $dl_\beta$ is the $\beta$ component of an infinitesimal portion of the path along $C_{ij}$.  
We stress that ${\bf f}_{ij}$ is the force between two molecules, i.e., the quantity that is responsible for the acceleration of the centres of mass of the molecules.  We consider neither torques nor forces between atoms on the same molecule nor forces of contstraint~\cite{tildesley1995}. 
For TIP4P/2005, ${\bf f}_{ij}$ is obtained by summing over all of the interactions between charge and Lennard-Jones sites on molecule $i$ and those on molecule $j$.
The freedom in choosing $C_{ij}$ renders the definition of the microscopic pressure non-unique.  Ikeshoji et al~\cite{FURUHOL} follows the convention of defining $C_{ij}$ to be a straight line segment connecting the centres of mass of molecules $i$ and $j$, consistent with the Irving-Kirkwood definition of the pressure tensor~\cite{kirkwood}.  As we comment below, this simple and intuitive choice of $C_{ij}$ leads to divergences in $P_T(r)$ that coarse-graining eliminates.

The coarse-graining procedure amounts to carrying out an integration of  Eq.~\ref{PKP} over ${\bf R}$ within a spherical shell of radius $r$, thickness $\Delta r$ and volume $\tilde V = 4 \pi [ R_{\rm out}^3 - R_{\rm in}^3 ]/3$, with $R_{\rm out}=r+\Delta r/2$ and $R_{\rm in}=r-\Delta r/2$.  We set $\Delta r=0.05$~nm. The coarse-grained pressure $P_{\alpha\beta}(r)$ is given by, 
\begin{equation}
P_{\alpha\beta}(r) = \frac{1}{\tilde V}\int_{\tilde V} P^\prime_{\alpha\beta}({\bf R})d{\bf R} = \left<P_{c,\alpha\beta}\right>+\left<P_{k,\alpha\beta}\right>.
\label{PKPV}
\end{equation}
The kinetic part is still calculated from the density, but now averaged over $\tilde V$.  The configurational part maintains the same form as before,  
\begin{eqnarray}
P_{c,\alpha\beta} &=& \frac{1}{2}\sum_i\sum_{j\ne i}P_{ij,\alpha\beta},\label{pconfV}
\end{eqnarray}
but now the coarse-grained contribution to the pressure from an interaction between a pair of molecules is given by,
\begin{eqnarray}
P_{ij,\alpha\beta} &=& \frac{1}{\tilde V} \int_{\tilde V}\int_{C_{ij}} f_{ij,\alpha}\,\delta({\bf R-l})dl_\beta d{\bf R},\nonumber \\
&=& \frac{1}{\tilde V} \int_{C_{ij}}f_{ij,\alpha}\left[\int_{\tilde V} \delta({\bf R-l})d{\bf R}\right]dl_\beta,\nonumber \\
&=& \frac{1}{\tilde V} \int_{C_{ij}\in \tilde V}f_{ij,\alpha}dl_\beta,\label{pijV}
\end{eqnarray}
where the force between molecules $i$ and $j$ contributes to the pressure in $\tilde V$ only along the parts of $C_{ij}$ that are in $\tilde V$. Regardless of the location of $i$ and $j$, i.e., whether they are in $\tilde V$ or not, as long as the line between them passes through $\tilde V$, their interaction contributes to the pressure.  
%

To determine the part of $C_{ij}$ that contributes to the pressure in $\tilde V$, one first uses a parametric expression for ${\bf l}(\lambda)$ that defines points located on $C_{ij}$,
\begin{equation}
{\bf l}(\lambda)={\bf r}_i + \lambda {\bf r}_{ij},
\label{llamb}
\end{equation}
where ${\bf r}_{ij} = {\bf r}_{j} - {\bf r}_{i}$, i.e. the vector pointing from $i$ to $j$.  (For repulsion, ${\bf f}_{ij}$ points approximately along ${\bf r}_{ij}$.)  Points on $C_{\ij}$ correspond to $\lambda \in [0,1]$.

Fig.~\ref{cgcases} shows a sketch of all possible contributions from molecular pair interactions to $\tilde V$ in the coare-grained method. The contributions from $C_{ij}$ that contribute to the pressure in $\tilde V$ are portions of lines between arrows, while the line between small filled circles is the line segment connecting particles $i$ and $j$.
For a given line, the portion between arrows corresponds to $\lambda_a\le\lambda\le\lambda_b$, with $a$ and $b$ labelling entry and exit points. If the line intersects $\tilde V$ over two segments (yielding two contributions to the pressure), there is a second set of entry and exit points that define the segment $\lambda_a^\prime\le\lambda\le\lambda_b^\prime$.
If ${\bf r}_i\in \tilde V$ then $\lambda_a=0$, while if ${\bf r}_j\in \tilde V$ then $\lambda_b$ (or $\lambda_b^\prime$ if it exists) $=1$.

A  precise determination of relevant  intersections between the line ${\bf l}(\lambda)$ and the spheres bounding $\tilde V$ requires solving the equation,
\begin{eqnarray}
{\bf l(\lambda)}\cdot {\bf l(\lambda)} &=& r_i^2 + \lambda \,2 {\bf r}_i\cdot{\bf r}_{ij} + \lambda^2 r_{ij}^2 = R_{\rm out}^2, \label{eq:l2}
\end{eqnarray}
and a similar one for $R_{\rm in}$.  The magnitudes of ${\bf r}_i$ and ${\bf r}_{ij}$ are $r_i$ and $r_{ij}$, respectively. The solutions to these quadratic equations are,
\begin{equation}\label{lambdasol}
\lambda^{\rm in/out}_{\pm} = -\frac{       {\bf r}_i\cdot{\bf r}_{ij}   }    {   r_{ij}^2 } \pm \frac{1}{r_{ij}^2 }\sqrt{D_{\rm in/out}},
\end{equation}
where the discriminants are given by,
\begin{equation}
D_{\rm in/out} =      \left(  {\bf r}_i\cdot{\bf r}_{ij} \right)^2    -  r_{ij}^2  \left(r_i^2 - R_{\rm in/out}^2\right).
\end{equation}
If $D_{\rm out} < 0$, there are no intersections and the pair interaction gives no contribution to the pressure in $\tilde V$.  All of the possible cases for solution sets yielding pressure contributions and the resulting limits of integration are given in Table~\ref{tablambda}.

\begin{table}
\begin{tabular}{ | c | c | c | c | c | c | c | c | c | c |}
\hline
$D_{\rm in}$  & $\lambda^{\rm in}_{-}$ & $\lambda^{\rm in}_{+}$ & $\lambda^{\rm out}_{-}$ & $\lambda^{\rm out}_{+}$ &
$\lambda_a$ & $\lambda_b$ & $\lambda_a^\prime$ & $\lambda_b^\prime$ & Case \\
\hline
$<0$ &  &  & [0,1] & [0,1] & $\lambda^{\rm out}_{-}$ & $\lambda^{\rm out}_{+}$ & & & C.1 \\
\hline
$>0$& $<0$ & $<0$ & $<0$ & $>1$ & 0 & 1 &  &  & C.2 \\
$>0$& $>1$ & $>1$ & $<0$ & $>1$ & 0 & 1 &  &  & C.2 \\
\hline
$>0$& [0,1] & [0,1] & [0,1] & [0,1] & $\lambda^{\rm out}_{-}$ & $\lambda^{\rm in}_{-}$ & $\lambda^{\rm in}_{+}$ & $\lambda^{\rm out}_{+}$ & C.3 \\
\hline
$>0$ & [0,1] & [0,1] & $<0$ & [0,1] & 0 & $\lambda^{\rm in}_{-}$ & $\lambda^{\rm in}_{+}$ & $\lambda^{\rm out}_{+}$ & C.4 \\
$>0$ & [0,1] & [0,1] & [0,1]  & $>1$ & $\lambda^{\rm out}_{-}$ & $\lambda^{\rm in}_{-}$ & $\lambda^{\rm in}_{+}$ & 1 & C.4 \\
\hline
$>0$ & $<0$ & [0,1] &  $<0$ & [0,1] & $\lambda^{\rm in}_{+}$ & $\lambda^{\rm out}_{+}$ & & & C.5 \\
$>0$ & [0,1] & $>1$ & [0,1] & $>1$ & $\lambda^{\rm out}_{-}$ & $\lambda^{\rm in}_{-}$ & & & C.5 \\
\hline
$>0$ & $<0$ & [0,1] & $<0$ & $>1$ & $\lambda^{\rm in}_{+}$ & 1 & & & C.6 \\
$>0$ & [0,1] & $>1$ & $<0$ & $>1$ & 0 & $\lambda^{\rm in}_{-}$ &  &  & C.6 \\
\hline
$>0$ & $<0$ & $<0$ & $<0$ & [0,1] & 0 & $\lambda^{\rm out}_{+}$ &  &  & C.7 \\
$>0$ & $>1$ & $>1$ & [0,1] & $>1$ & $\lambda^{\rm out}_{-}$ & 1 &  &  & C.7 \\
\hline
$>0$ & [0,1] & [0,1] & $<0$ & $>1$ & 0 & $\lambda^{\rm in}_{-}$ & $\lambda^{\rm in}_{+}$ & 1 & C.8 \\
\hline
$<0$ &  &  & $<0$ & [0,1] & 0 & $\lambda^{\rm out}_{+}$ &  &  & C.9 \\
$<0$ &  &  & [0,1] & $>1$ & $\lambda^{\rm out}_{-}$ & 1 &  &  & C.9 \\
\hline
$<0$ &  &  & $<0$ & $>1$ & 0 & 1 &  &  &  C.10\\
\hline
\end{tabular}
\caption{List of all 16 solution sets of Eq.~\ref{lambdasol} that contribute to  Eq.~\ref{pijVlamb} and the resulting limits of integration.  In all cases $D_{\rm out}>0$.  Entries in the rightmost column refer to curve labels in Fig.~\ref{cgcases}.}
\label{tablambda}
\end{table}

Having determined all intersections and limits on  our integration variable $\lambda$, Eq.~\ref{pijV} becomes,
\begin{equation}
\begin{split}
P_{ij,\alpha\beta} =  \frac{1}{\tilde V} \left[ \int_{\lambda_a}^{\lambda_b}({\bf f}_{ij}\cdot {\bf e}_\alpha)({\bf r}_{ij}\cdot {\bf e}_\beta)d\lambda  \right. \\ +
\left. \int_{\lambda_a^\prime}^{\lambda_b^\prime}({\bf f}_{ij}\cdot {\bf e}_\alpha)({\bf r}_{ij}\cdot {\bf e}_\beta)d\lambda  
\right],\label{pijVlamb}
\end{split}
\end{equation}
where the integrand is expressed in terms of the unit vectors ${\bf e}_r$, ${\bf e}_\theta$, and ${\bf e}_\phi$. 
Note that if there is only one portion of $C_{ij}$ intersecting $\tilde V$, then the second integral in Eq.~\ref{pijVlamb} (with limits $\lambda_a^\prime$ and  $\lambda_b^\prime$) is absent.
These unit vectors are not constant as ${\bf l}(\lambda)$ moves along $C_{ij}$, and the unit vectors in Cartesian coordinates are,
\begin{eqnarray}
{\bf e}_r &=& \left\{\frac{l_x}{l},\frac{l_y}{l},\frac{l_z}{l}\right\}, \label{unitv} \\
{\bf e}_\theta &=& \left\{\frac{l_xl_z}{l(l_x^2+l_y^2)^{1/2}},\frac{l_yl_z}{l(l_x^2+l_y^2)^{1/2}},\frac{-(l_x^2+l_y^2)^{1/2}}{l}\right\},\nonumber \\
{\bf e}_\phi &=& \left\{\frac{-l_y}{(l_x^2+l_y^2)^{1/2}},\frac{l_x}{(l_x^2+l_y^2)^{1/2}},0\right\},\nonumber
\end{eqnarray}
where $\phi$ is the azimuth angle in the $xy$-plane, $\theta$ is the angle between {\bf l} and the z-axis, $l=\left|\,{\bf l}\,\right|$, and $l_\alpha$ is the $\alpha$ component of ${\bf l}$.  

The $P_{ij}$  tensor can be written in terms of two components, normal and tangential. These components are obtained from Eq.~\ref{pijVlamb} using Eqs.~\ref{llamb} and~\ref{unitv}.  The contribution from the interaction between molecules $i$ and $j$ to the normal component is given by
\begin{eqnarray}
P_{ij,N} &\equiv& P_{ij,rr} =  \frac{1}{\tilde V} \int_{\lambda_a}^{\lambda_b}({\bf f}_{ij}\cdot {\bf e}_r)({\bf r}_{ij}\cdot {\bf e}_r)d\lambda  \label{lambdaint1} \\
              &=&  \frac{1}{\tilde V}  \int_{\lambda_a}^{\lambda_b}\frac{a_n+b_n \lambda + c_n \lambda^2}{d_n+e_n \lambda + f_n \lambda^2}d\lambda \label{lambdaint2} \\
              &=& \frac{1}{\tilde V} \left\{ \Sigma_N(\lambda)\Bigr|_ {\lambda_a}^{\lambda_b} -  \Sigma_N(\lambda)\Bigr|_ {\lambda_a^\prime}^{\lambda_b^\prime} \right\},
\end{eqnarray}
where we omit in Eqs.~\ref{lambdaint1} and~\ref{lambdaint2} the second integral simply for brevity, and, 
\begin{eqnarray}
a_n&=& \left(   {\bf r}_i \cdot {\bf f}_{ij}   \right)  \left(   {\bf r}_i \cdot {\bf r}_{ij}    \right) \\
b_n &=& \left(   {\bf r}_{ij} \cdot {\bf f}_{ij}   \right) \left(   {\bf r}_i \cdot {\bf r}_{ij}    \right) +  \left(   {\bf r}_i \cdot {\bf f}_{ij}   \right)  \left(   {\bf r}_{ij} \cdot {\bf r}_{ij}    \right)\nonumber \\
c_n &=& \left(   {\bf r}_{ij} \cdot {\bf f}_{ij}   \right) \left(   {\bf r}_{ij} \cdot {\bf r}_{ij}    \right) \nonumber  \\
d_n &=&     {\bf r}_i \cdot {\bf r}_{i}   \nonumber  \\
e_n &=& 2 \,   {\bf r}_i \cdot {\bf r}_{ij} \nonumber \\
f_n &=&   {\bf r}_{ij} \cdot {\bf r}_{ij}  \nonumber \\
\Sigma_N(\lambda) &=& \frac{1}{2 f_n^2} \left \{   2 c_n f_n \lambda   +  \left (b_n f_n - c_n e_n\right)    \right.   \\ 
 & \times& \ln{\left[ d_n + e_n \lambda + f_n \lambda^2 \right]} \nonumber \\
    & +& \frac{2}{\sqrt{4d_n f_n -e_n^2}}  \arctan{\left[ \frac{e_n + 2 f_n \lambda}{\sqrt{4d_n f_n -e_n^2}} \right]} \nonumber \\
   &\times& \left.  \left( f_n (2 a_n f_n - b_n e_n) + c_n (e_n^2 - 2 d_n f_n)      \right) \right\},   \nonumber
\end{eqnarray}
while the tangential component is given by,
\begin{eqnarray}
P_{ij,T} &\equiv& P_{ij,\phi\phi} =  \frac{1}{\tilde V} \int_{\lambda_a}^{\lambda_b}({\bf f}_{ij}\cdot {\bf e}_\phi)({\bf r}_{ij}\cdot {\bf e}_\phi)d\lambda  \label{lambdaint3} \\
              &=&  \frac{c_t}{\tilde V}  \int_{\lambda_a}^{\lambda_b} \frac{a_t+b_t \lambda}{d_t+e_t \lambda + f_t \lambda^2}d\lambda \label{lambdaint4} \\
              &=& \frac{c_t}{\tilde V} \left\{ \Sigma_T(\lambda)\Bigr|_ {\lambda_a}^{\lambda_b} -  \Sigma_T(\lambda)\Bigr|_ {\lambda_a^\prime}^{\lambda_b^\prime} \right\},
\end{eqnarray}
where we omit in Eqs.~\ref{lambdaint3} and~\ref{lambdaint4} the second integral for brevity, and, 
\begin{eqnarray}
a_t &=& r_{i,x} \, f_{ij,y} -  r_{i,y} \, f_{ij,x} \\
b_t &=&  r_{ij,x} \, f_{ij,y}  - r_{ij,y} \, f_{ij,x} \nonumber \\
c_t &=& r_{i,x} \, r_{ij,y}  - r_{i,y} \, r_{ij,x} \nonumber \\
d_t &=&     r_{i,x}^2 + r_{i,y}^2  \nonumber  \\
e_t &=& 2 \, \left(  r_{i,x} \, r_{ij,x} + r_{i,y} \, r_{ij,y} \right) \nonumber  \\
f_t &=&   r_{ij,x}^2 +r_{ij,y}^2 \nonumber  \\
\Sigma_T(\lambda) &=& \frac{b_t \ln{\left[ d_t +e_t \lambda + f_t \lambda^2 \right]}}{2 f_t} \label{eq:sigmaT}\\ 
 & + & \arctan{\left[  \frac{e_t + 2 f_t \lambda}{\sqrt{4 d_t f_t - e_t^2}} \right] } \frac{(2 a_t f_t - b_t e_t) }{f_t \sqrt{4 d_t f_t - e_t^2}} \nonumber 
\end{eqnarray}

Having assembled all the pieces required to calculate the coarse-grained pressure tensor components, we now  report on the following radial quantities related to the pressure (see Eq.~\ref{PKPV}):
\begin{eqnarray}
P_N(r) &=& \left< P_{c,rr} \right> + k_B T \rho(r) \\
\bar{P}_{c,N}(r) &=& \left< P_{c,rr} \right> \\
P_T(r) &=&   \left< P_{c,\phi\phi} \right> + k_B T \rho(r)  \\
\bar{P}_{c,T}(r) &=& \left< P_{c,\phi\phi} \right> \\
P(r) &\equiv& \frac{1}{3} P_N(r) + \frac{2}{3} P_T(r)
\end{eqnarray}
where $\rho(r)$ is the average number density of molecules in $\tilde V$  as determined from molecular centres of mass and $P(r)$, the mean (or isotropic) pressure, is one third the trace of the pressure tensor.  As noted in Ref.~\cite{FURUHOL}, the tangential component may be calculated from 
$P_{ij,\theta\theta}$.  However, the analytic expression for the resulting antiderivative is very cumbersome.

\subsection{Comment on calculating $P_T(r)$}

\begin{figure}
 \centering
 \includegraphics[width=0.9\columnwidth]{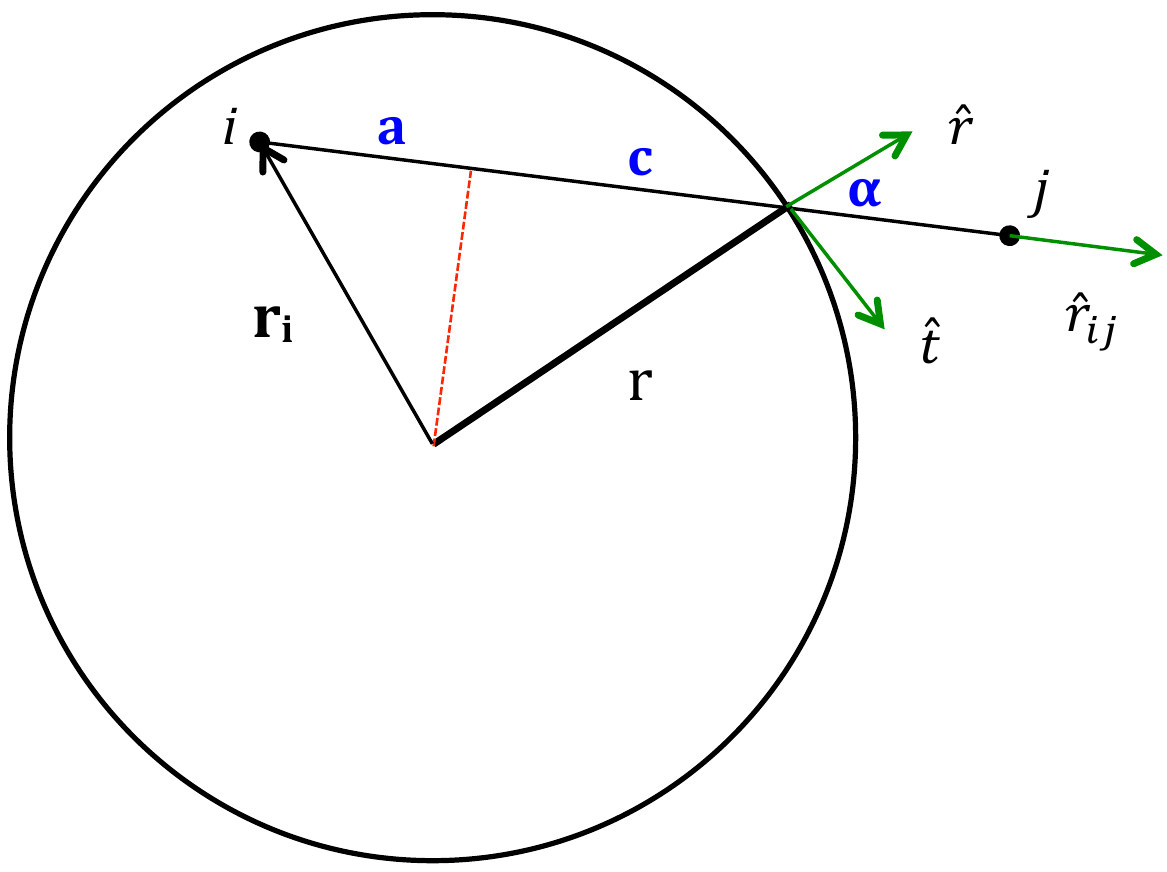}
 \caption{
 A sketch of the geometry for a sample calculation of the transverse pressure component at a radius $r$.  The straight line contour intersects the sphere when $\lambda=\lambda_0$ (see Eq.~\ref{llamb}), at which point $a+c = \lambda_0 r_{ij}$.  Here, the force between $i$ and $j$ is taken to be radial, and forms an angle $\alpha$ with $\hat{r}$, with $\cos\alpha = c/r$ and $a= - {\bf r}_i \cdot \hat{r}_{ij}$.
 }
 \label{pt}
 \end{figure}

Without coarse-graining, the transverse component of the pressure tensor is calculated from the first of two equivalent equations relating pressure components derived from the condition
of  mechanical stability~\cite{thomp,FURUHOL},
\begin{eqnarray}
P_T(r) &=& P_N(r) + \frac{r}{2}\frac{d P_N(r)}{dr} \label{eq:pncheck}\\
P_N(r) &=& \frac{2}{r^2}\int_0^r P_T(r^\prime) r^\prime dr^\prime, \label{eq:ptcheck}
\end{eqnarray}
rather than directly from configurations on account of divergences occurring in Eq.~\ref{pij}.  
(We note that Eqs.~\ref{eq:pncheck} and~\ref{eq:ptcheck} are valid regardless of whether the quantities are coarse-grained or not.)
To illustrate this, let us use Eq.~\ref{pij} in the context of calculating the transverse pressure component over a sphere (not a spherical shell) of radius $r$ and assume for simplicity, for the purposes of this illustration only, that ${\bf f}_{ij} = f_{ij} {\hat r}_{ij}$, with $f_{ij}$ a scalar and the unit vector is the one derived from ${\bf r}_{ij}$, i.e. that the force is central - acting along the line joing the particles.  Our setup for this illustration is shown in Fig.~\ref{pt}, where we take the transverse direction to be in the plane of ${\bf r}_{ij}$ and $\hat{r}$, the radial unit vector at the point of intersection of $C_{ij}$ with the sphere, at which point $\lambda=\lambda_0$. As we are now considering the contribution to the pressure over the spherical surface, Eq.~\ref{pij} becomes,
\begin{eqnarray}
P^\prime_{ij,T}(r) &=& \frac{1}{2} \frac{1}{4 \pi r^2} \int_{C_{ij}} \left( {\bf f}_{ij}\cdot \hat{t} \, \right)  \left( d{\bf l}\cdot \hat{t}\, \right) \delta(r-l\,) \\
      &=& \frac{1}{8 \pi r^2} \,f_{ij}  \, r_{ij}  \sin^2\alpha\, \int_0^1 d\lambda \,\delta \big(r-l(\lambda)\big) \\
      &=&  \frac{1}{8 \pi r^2} \,f_{ij}  \, r_{ij}  \sin^2\alpha\, \int_0^1 d\lambda \, \frac{\delta(\lambda - \lambda_0)}{\left| l^\prime(\lambda_0) \right|} \\
      &=&  \frac{1}{8 \pi r^2} \,f_{ij} \, \frac{\sin^2\alpha}{\cos\alpha} \label{eq:pprimet},
\end{eqnarray}
where the extra factor of $\frac{1}{2}$ comes from $\hat{t}$ having both $\theta$ and $\phi$ components and, with the help of Eq.~\ref{eq:l2} and the geometrical arrangement shown in Fig.~\ref{pt}, it can be shown that
$l^\prime(\lambda_0)= r_{ij} \cos\alpha$.  Eq.~\ref{eq:pprimet} appears in Ref.~\cite{FURUHOL} as Eq.~12, which is itself referenced from~\cite{HSP}.  The cosine in the denominator causes a divergence when $\cos\alpha=0$, i.e., when the $C_{ij}$ becomes tangent to the sphere.  Attempts to use Eq.~\ref{eq:pprimet} to calculate the transverse pressure illustrate the problem, which is formally absent in the coarse-graining method because of the order in which the integration is carried out in obtaining Eq.~\ref{pijV}.
 %


\subsection{Obtaining the local pressure from the potential energy}
Ikeshoji et al~\cite{FURUHOL} also discusses the method of determining the pressure tensor in $\tilde V$ by using the virial expression for the bulk pressure, 
but only considering particle interactions for which at least one of the particles is in  $\tilde V$.  While this intuitive approach is only a low-order approximation~\cite{Daivis1995}, the authors demonstrate for a planar geometry that it fails to respect mechanical equilibrium (Eqs.~\ref{eq:pncheck} and~\ref{eq:ptcheck}) only at the interface.

\begin{figure}[!htb]
 \centering
 \includegraphics[width=0.7\columnwidth]{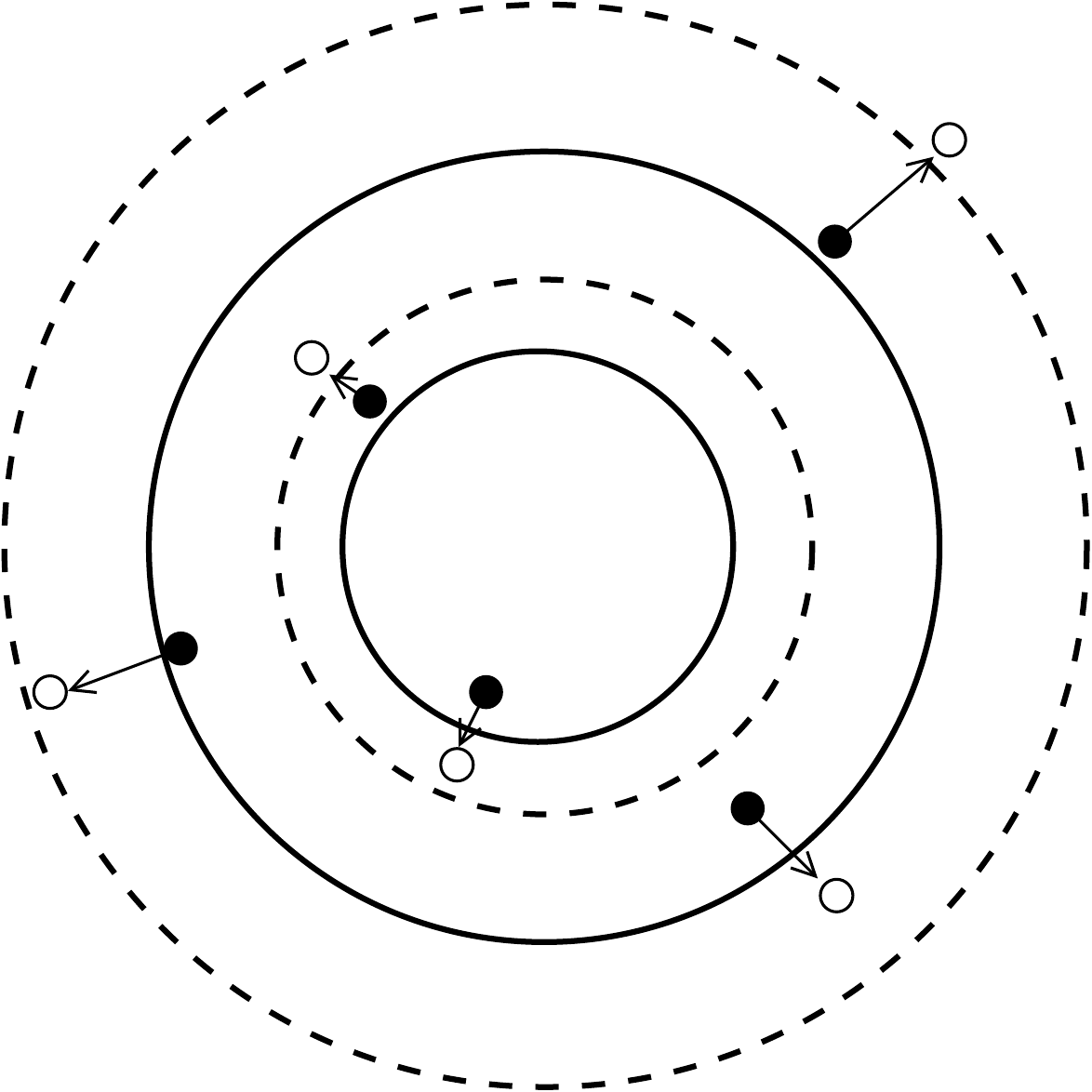}
 \caption{
 A sketch for the calculation of the derivative of the local potential energy $U(r)$ with respect to volume.  All particles coordinates are rescaled isotropically according to ${\bf r} \rightarrow (1+\alpha) {\bf r}$ (filled to open circles), resulting in a commensurate change in spherical shell volume $\tilde V$ (solid lines) to $(1+\alpha)^3 \tilde V$ (dashed lines).
 }
 \label{fig:fran}
 \end{figure}

In the same spirit,  we define an expression inspired by the thermodynamic meaning of pressure in the bulk, 
\begin{eqnarray}
P_U(r) \equiv  \rho(r) k_B T - \left< \frac{dU(r)}{d\tilde V} \right>_{T,N},
\end{eqnarray}
where the derivative is calculated in the following way (see Fig.~\ref{fig:fran}).   For a given nanodroplet configuration, 
all molecular centres of mass are  isotropically expanded according to 
${\bf r}_{CM,i}^+ \rightarrow (1+\alpha_+) {\bf r}_{CM,i}$,
and in this rescaled system we calculate the binding energy 
$u_i^+= \sum_{j\ne i} u_{ij}$ for each molecule $i$ originally in $\tilde V$, where $u_{ij}$ is the interaction energy between molecules $i$ and $j$.
The rescaled shell volume is $\tilde V_+ = (1+\alpha_+)^3 \tilde V$, and the potential energy associated with the rescaled shell is 
$U_+=\frac{1}{2} \sum_{i \in \tilde V_+} u_i^+$.  To use the centred difference scheme to approximate the derivative,
\begin{eqnarray}
\frac{dU(r)}{d\tilde V} \approx \frac{U_+ - U_-}{\tilde V_+ - \tilde V_-},
\end{eqnarray}
we similarly rescale the molecular centres of mass according to  ${\bf r}_{CM,i}^- \rightarrow (1+\alpha_-) {\bf r}_{CM,i}$ to obtain $U_-$ and $\tilde V_-$.
We use $\alpha_+=10^{-4}$, and then to ensure that  $\tilde V_+ - \tilde V =  \tilde V - \tilde V_-$, we use 
$\alpha_- = \left[ 2 - (1+\alpha_+)^3 \right]^{1/3} -1$ (approximately equal to $-\alpha_+$).  Note that the same particles are in $\tilde V$, $\tilde V_+$ and $\tilde V_-$ and that the same molecular pairs are used to calculate $U_+$ and $U_-$.  This derivative is then averaged over nanodroplet configurations.

\section{Results}\label{sec:res}

\subsection{Radial pressure profiles}

 \begin{figure}[!htb]
 \centerline{\includegraphics[scale=0.35]{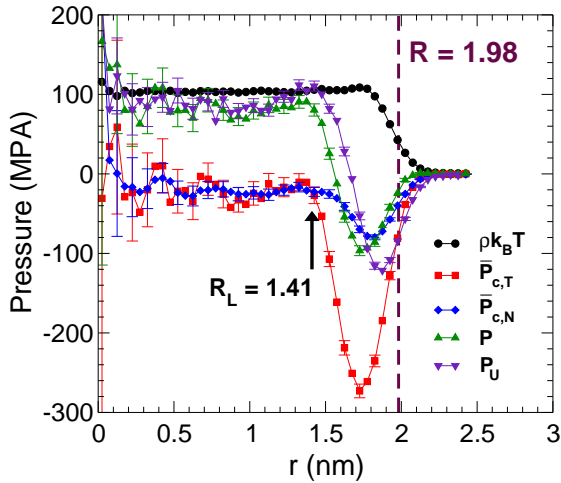}}
 \caption{
 Pressure as a function of radial distance from the centre of a nanodroplet of size $N=1100$ at $T=220$~K.  The radial extent of the droplet is estimated by $R=\sqrt{5/3}R_g=1.98$~nm, 
 while the configurational contributions to the tangential and radial pressures are approximately equal below the point of crossing at $R_L = 1.41$~nm. 
 }
 \label{fig:pcomp}
 \end{figure}

In Fig.~\ref{fig:pcomp} we plot various pressure contributions for a nanodroplet of size $N=1100$.
The radial density is proportional to the ideal gas term (black circles), which for this state point accounts for most of the roughly 100~MPa of pressure in the interior of the nanodroplet.  There is a small maximum in the density at or near the surface (at $r\approx 1.75$~nm,) where the configurational contributions to the normal [$\bar{P}_{c,N}(r)$ - blue diamonds] and tangential [$\bar{P}_{c,T}(r)$ - red squares] components of the pressure are maximally negative.  Despite the large negative values near the surface, $\bar{P}_{c,N}(r)$ and $\bar{P}_{c,T}(r)$ become indistinguishable from each other within the precision of our simulations below  $R_L \approx 1.41$~nm, indicating that the pressure tensor is isotropic within this radius.

We note that an accurate determination of the centre of mass of the cluster is vital for determining all the radial quantities. It is thus important to exclude gas-like molecules when calculating the centre of mass.  When calculating the pressure all particles in the system are used.  However, the vapour pressure at $T=220~K$ is nearly zero.  For example, a search of the $N=1100$ configurations sampled, using the definition that a gas-like molecule has two or fewer neighbours within $r_n = 0.63$~nm, found no such molecules.  A cluster search employing the definition that two molecules within $r_n=0.35$ belong to the same cluster yields the same result~\cite{sevick1988}.

Notwithstanding the progressively larger error bars as $r \rightarrow 0$, there appear to be oscillations within both $\bar{P}_{c,N}(r)$ and $\bar{P}_{c,T}(r)$ that may correlate with small oscillations in $\rho(r)$.  However, given the precision of our calculations,  we can do no better than to assume that $\bar{P}_{c,N}(r)$ and $\bar{P}_{c,T}(r)$ are both equal to the same constant below $R_L$.

\begin{figure}[!htb]
 \centerline{\includegraphics[scale=0.35]{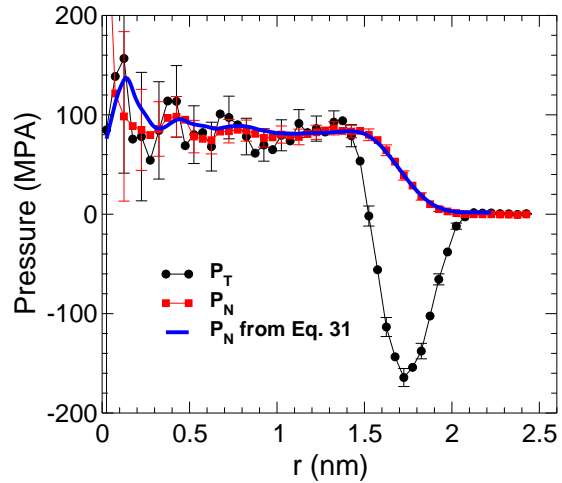}}
 \caption{
 Consistency check on the calculation of $P_N(r)$ and $P_T(r)$ for $N=1100$ and $T=220$~K.
 }
 \label{fig:consistency}
 \end{figure}

As a consistency check on our results, we verify that our calculated pressure components satisfy mechanical equilibrium by 
using Eq.~\ref{eq:ptcheck} to recover $P_N(r)$ from $P_T(r)$.  We use Eq.~\ref{eq:ptcheck} instead of Eq.~\ref{eq:pncheck} since 
numerical integration reduces noise.  
In Fig.~\ref{fig:consistency} we plot both $P_N(r)$ calculated directly from the droplets and as calculated from Eq.~\ref{eq:ptcheck}.  We see that the two curves are the same within error, even though Eq.~\ref{eq:ptcheck} yields a curve with less pronounced oscillatory behaviour.  A global estimate of the numerical integration error can be taken to be the difference between Eq.~\ref{eq:ptcheck} and $P_N(r)$ where the latter  decays to zero.

$P_U(r)$ for the same state point is shown in Fig.~\ref{fig:pcomp}, where it agrees, to within error, with $P(r)$ in the interior of the droplet where the pressure is constant with $r$.   At the interface, there is a significant difference, in which $P_U(r)$ exaggerates the extremal values of $P(r)$, and  shows a positive pressure peak near the surface.  Despite this exaggeration near the surface, $P_U(r)$ shows none of the apparent oscillations seen in $P(r)$.

As this method only relies on the potential energy, it is comparatively a rather straightforward calculation, and so may be of use when interactions are complex and precise determination of the properties near the interface is not required.  Furthermore, that the two methods agree within the interior provides a useful check on the results for $P(r)$.

%
%

\subsection{Laplace pressure relation}

To test Eq.~\ref{eq:laplace}, and noting that the vapour pressure is so small compared to the interior pressure of the nanodroplets,
we simply define $P_L$ to be the average of $P(r)$ from $r_{\rm min}=0.025$ (our first data point) to $R_L$, the radial distance to which
the pressure tensor is isotropic, i.e., below which point $P_T(r)$ and $P_N(r)$ are indistinguishable:
\begin{eqnarray}
P_L \equiv  \frac{3}{4 \pi \left( R_L^3 - r_{\rm min}^3 \right)} \int_{r_{\rm min}}^{R_L} 4 \pi r^2 P(r) dr.
\end{eqnarray}
Operationally, we take $R_L$ to be the
first crossing of $P_{c,T}(r)$ and $P_{c,N}(r)$ as $r$ decreases below the location of the minimum in $P_{c,T}(r)$.
As a measure of the radius of the droplet,  treating the nandroplets as spheres of uniform density, we choose 
$R=\sqrt{5/3} R_g$, where $R_g$ is the radius of gyration.

 \begin{figure}[!htb]
 \centerline{\includegraphics[scale=0.35]{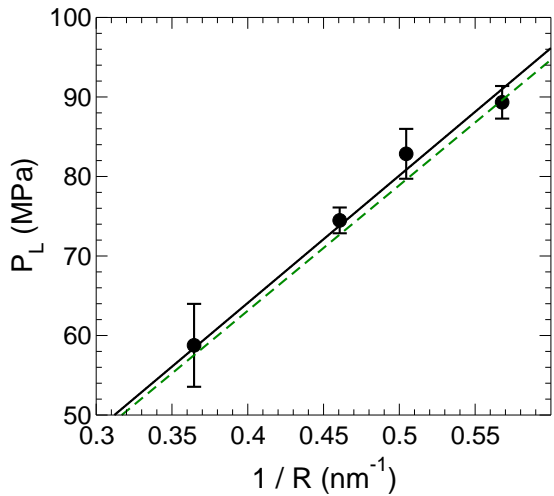}}
 \caption{
 Test of the Laplace pressure relation.  Plotted is the average isotropic pressure from the interior of nanodroplets as a function of
 $1/R$, where $R=\sqrt{5/3} R_g$.  Solid line is the result of a one-parameter least-squares fit, $P_L=2(80.1)/R$. 
 The dashed line uses an estimate of $\gamma=78.9$~mN/m for a planar interface at 220~K~\cite{Vega2007}.
 }
 \label{fig:laplace}
 \end{figure}

In Fig.~\ref{fig:laplace} we plot $P_L$ as a function of $1/R$.  We fit the data to $2\gamma_{\rm fit} /R$ and find $\gamma_{\rm fit}=80.1$.  This estimate of $\gamma$ agrees well with the value
$\gamma=78.9$~mN/m obtained using Eq.~6 in Ref.~\cite{Vega2007}; the dashed line in Fig.~\ref{fig:laplace} shows $2\gamma /R$ using this value of $\gamma$.


\section{Discussion and conclusions}\label{sec:dis}

Calculating the local pressure is a non-trivial task and requires good averaging because of significant statistical fluctuations, particularly at small radial distances.  We note the discrepancy between our results and the early work on ST2 water clusters of Brodskaya et al.~\cite{brodskaya1987,brodskaya1993}.  They reported a significant drop in the pressure, even to significantly negative values, towards the centre of the droplet.  While the droplet sizes they investigated were smaller and at higher $T$, we speculate that this unexpected result may have arisen from an imprecise determination of the centre of mass or even from sample bias since these early simulations had much shorter run times.  A given configuration may have an extremely large (positive or negative) value of $P(r\rightarrow0)$, depending on whether there is a high or low density fluctuation at the centre, which can be considerable given the small number of particles there.  As a general remark, statistics for larger $r$ are not only better because of the greater volume over which the average is determined, but because mobility is likely greater the closer a layer is to the surface.
However, in the present study we have not excluded the possibility that for smaller droplets, such as those studied in Refs.~\cite{brodskaya1987,brodskaya1993}, there exists an effect that reduces the pressure at the centre.

It is important to directly calculate the pressure instead of relying only on the local density and the known bulk equation of state, even when done as elegantly as in a recent test of the Young-Laplace equation for the SPC/E model by pressuring water through a nanopore~\cite{Liu2016}.  We already see a dense region near the surface of the nanodroplet, where the pressure is negative.  Clearly, the water in this layer does not follow the bulk equation of state.  Further, subtle finite size effects on structure, as noted already in regard to nucleation~\cite{Debenedetti2017}, may affect local pressure more than local density.  Thus, water in sufficiently small nanodroplets may not follow the bulk equation of state.  

Whether or not droplet interiors represent bulk water also depends on how deeply surface effects propagate inside.  At $T=220$~K, we see, coming in from large $r$, that the density rises from zero to a local maximum [where $P(r)$ is most negative] in about 0.3~nm (see black curve with circles in Fig.~\ref{fig:pcomp}).  Another 0.4~nm further inside and $P_N(r)$ and $P_T(r)$ become indistinguishable within uncertainty.  This non-bulk-like region is 0.7~nm thick and encompasses approximately two molecular layers.  This estimate of the size of non-bulk-like region is somewhat smaller than pointed out in Ref.~\cite{Debenedetti2017}, for which there is also observed a local maximum in the stress before quickly tending to a constant at smaller $r$.   However, in our case the interior is at a high pressure and the definition of the local stress used in~\cite{Debenedetti2017} differs from that of the pressure.  We note that $P_U(r)$ also produces a peak near $R_L$, and would thus also produce a larger estimate of the extent of the non-bulk-like region.  This should not be an issue if one is in search of a conservative estimate of what is perhaps bulk-like.


Eq.~\ref{eq:laplace} formally models a droplet with a sharp interface at $R=R_s$, at the so-called surface of tension, that separates interior and exterior fluids with isotropic and homogeneous pressures, and $\Delta P$ refers to the difference between these fluid pressures.   For our droplets, the pressure tensor components become equal and constant with $r$ near the centre (and hence bulk-like), and so we identify $\Delta P$ with $P_L$ obtained from the pressure tensor. 
In using Eq.~\ref{eq:laplace} we approximate $R_s$ with $\sqrt{5/3} R_g$.  In a more systematic study aiming to quantify the curvature corrections to $\gamma$ (through the Tolman length $\delta$), the choice of dividing surface should be carefully considered. 
Nonetheless, our use of $R=\sqrt{5/3} R_g$ yields a $\gamma$ remarkably consistent with the expected planar value.  This may indicate that  curvature corrections to $\gamma$, and hence $\delta$ itself, are small.
Calculations for both Lennard-Jones~\cite{giessen2013} and TIP4P/2005~\cite{peters2013} yield small negative values of $\delta$, around -0.1$\sigma$ and $-0.05$~nm respectively, with the magnitude of $\delta$ decreasing with decreasing $T$ for TIP4P/2005~\cite{reguera2015}.
For a future study of smaller droplets, for which curvature effects may become more apparent, the pressure calculation presented here provides the means of  directly determining $\delta$ from simulation data,  as has been done for Lennard-Jones droplets~\cite{Blokhuis2009}.  In addition, density functional theory suggests that $\delta$ becomes positive for very small droplets, as implied by a decreasing $\gamma$ with $R_s$~\cite{ghosh2011}, and hence in the present study we may be in a droplet size regime where $\delta\approx 0$.


While working with forces between molecules and their centres of mass is more convenient compared to treating molecules as collections of atoms held rigidly by forces of constraint, there is another important advantage of our approach.  As recently pointed out by Sega et al.~\cite{sega2017},
when constraints are used and the kinetic energy tensor is calculated from atomic velocities, the kinetic energy tensor may become anisotropic at a liquid-vapour interface.  Failure to consider these anisotropies may, for example, lead to underestimates of $\gamma$ by approximately 15\% for a planar interface.
It is thus insufficient, when working with constraints, to only calculate the configuration contribution to the virial and assume an isotropic ideal gas contribution.  Velocities are thus required for the pressure calculation.  In contrast, we work with the velocities of the molecular centres of mass and intermolecular forces, thus avoiding these difficulties~\cite{tildesley1995}.  The molecular approach works essentially because the calculation of pressure stems from the calculation of the force, i.e., the rate of change of the linear momentum with time~\cite{Schofield}.  The validity of the molecular approach used here, where we assume an isotropic ideal gas contribution, is confirmed in Fig.~\ref{fig:consistency}, where $P_N(r)$ and $P_T(r)$ are shown to be consistent with mechanical stability.  If our ideal gas contribution were incorrect, mechanical stability would appear to be violated.  Regardless of the concerns raised by Sega et al.~\cite{sega2017}, our estimates for $P_L$ are made solely based on the behaviour of the pressure tensor in the interior of the droplets.  As a result,  anisotropy arising in the region of the surface does not affect our results for $P_L$.  

In summary, we have provided a detailed description of the calculation of the microscopic pressure for spherical droplets of molecular liquids, and checked the results by introducing an approximate energy-based method of calculating the microscopic isotropic pressure.  Our calculation paves the way for a detailed analysis of effects of the local pressure on nucleation, and for direct checks on whether the bulk equation of state remains valid in nanodroplet interiors.  For the size range studied, and at fairly deeply supercooled $T$, we find that $\gamma$ determined from a flat interface predicts the pressure in the interior of the nanodroplet quite well, despite significant surface features in the radial dependence of the pressure.


%




\section*{Acknowledgments}

We thank Richard K.~Bowles for enlightening discussions.
We thank Natural Sciences and Engineering Research Council (Canada) for funding.
Computational facilities are provided by ACENET, a member of Compute Canada and 
the regional high performance computing consortium for universities in Atlantic Canada. 
ACENET is funded by the Canada Foundation for Innovation (CFI), 
the Atlantic Canada Opportunities Agency (ACOA), 
and the provinces of Newfoundland and Labrador, Nova Scotia, and New Brunswick.

\end{document}